# Metal to Insulator Transition, Colossal Seebeck Coefficient and Large Violation of Wiedemann−Franz law in Nanoscale Granular Nickel

Vikash Sharma[1,2,*], Gunadhor Singh Okram[1,#] and Yung-Kang Kuo[3]

[*]vikash.sharma@tifr.res.in, [#]okram@csr.res.in

[1]UGC-DAE Consortium for Scientific Research, University Campus, Khandwa Road, Indore 452001, Madhya Pradesh, India.

[2]Department of Condensed Matter Physics & Materials Science, Tata Institute of Fundamental Research, Homi Bhabha Road, Mumbai-400005, India

[3]Department of Physics, National Dong-Hwa University, Hualien 97401, Taiwan

**ABSTRACT:** We report on the electrical and thermal transport properties of nickel nanoparticles with crystallite size from 23.1 ± 0.3 to 1.3 ± 0.3 nm. These nanoparticles show a systematic metal to insulator transition with the change in the conduction type from *n*- to *p*-type, colossal Seebeck coefficient of 1.87 ± 0.07 mVK$^{-1}$, and ultralow thermal conductivity of 0.52 ± 0.05 Wm$^{-1}$K$^{-1}$ at 300 K as the crystallite size drops. The electrical resistivity analysis reveals a dramatic change in the electronic excitation spectrum indicating the opening of an energy gap, and cotunneling and Coulomb blockade of the charge carriers. Seebeck coefficient shows transport energy degradation of charge carriers as transport level moves away from the Fermi level with decrease in crystallite size. The Lorenz number rising to about four orders of magnitude in the metallic regimes with decrease in crystallite size, showing a large violation of the Wiedemann−Franz law in these compacted nickel nanoparticles. Such an observation provides the compelling confirmation for unconventional quasiparticle dynamics where the transport of charge and heat is independent of each other. Therefore, such nanoparticles provide an intriguing platform to tune the charge and heat transport, which may be useful for thermoelectrics and heat dissipation in nanocrystal array-based electronics.

**INTRODUCTION**

Nanoscale granular metals or metallic nanoparticles (NPs) are considered to be the most suitable systems for exploring the novel electronic and thermal transport properties at nanoscale[1,2,3,4,5]. The unique properties of NP's core are combined with insulating matrix or surfactant/s as well as collective and correlation-driven effects between the NPs produces novel properties relative to bulk counterpart[2]. The transport properties of NPs mainly depend on competition of mean energy level spacing, Coulomb charging energy for a single particle and tunnel energy associated with the inter-particle coupling[6]. In densely packed NPs, transport of electrons can possible both through nearest neighbors hopping and cotunneling of electrons between adjacent NPs via a chain of intermediate virtual states[2]. However, in the limit of weak coupling between NPs, electrons are transported via phonon-assisted tunneling between NPs through the insulating matrix that disconnects the adjacent NPs, wherein single electron charging becomes significant resulting in Coulomb blockade behavior and the localization of electrons on individual particles[2,3,6].

In order to realize the interesting physical properties at nanoscale, manipulating the materials with controlled size, size-distribution, and shape is crucial. However, colloidal synthesis of NPs provide nearly precisely controlled platform for controlling the particle size, size distribution and shape using variety of surfactant/s[7]. In this context, transport properties of compacted nickel nanoparticles (NPs) of varying average sizes could be of special interest when the size is small and uniform. It is because they show many fascinating properties of natural nanolattice formation[8], structural phase transition in the smaller particle sizes[9], anomalous electrical transport[10–12], quantum size effect in heat capacity[13] and superparamagnetism[14] as the crystallite size changes.

In the present work, we show some remarkable properties of compacted monodispersed Ni NPs, with a systematic crystallite size control through the variation in the quantity of trioctylphosphine (TOP) in preparing the samples. For example, we established the large evolutionary violation of Wiedemann−Franz law (WFL) with the Lorenz number increasing to nearly four orders of magnitude compared to a universal Sommerfeld value $L_0 = \frac{\pi^2}{3}\left(\frac{k_B}{e}\right)^2 = $ 2.44x10$^{-8}$ WΩK$^{-2}$. The WFL defines as $L_o = \frac{\kappa_e}{\sigma T}$, where $\kappa_e$ is electronic part of thermal conductivity and $\sigma$ is electrical conductivity[15], is a basic property of Fermi liquid (FL) in which same quasiparticle that carries charge e, and also transport heat energy of the order of $K_B T$. Validity of WFL in a variety of materials has firmly established the quasiparticle picture i.e. the ground state of every metal investigated so far is an FL[16]. However, violation of the WFL (i.e. $L \neq L_0$[17]), has also been proposed theoretically[18–20] and found experimentally[21–25] in various materials including Luttinger liquids[18], VO$_2$ nanobeams[21], nanoscale Au and Pt[19], granular metal[20], Bi$_2$Te$_2$Se thin films[22], Au thin films[23], graphene[24] and quasi-one-dimensional (1D) Li$_{0.9}$Mo$_6$O$_{17}$ conductor[25]. Nonetheless, this violation has so far been mostly marginal with Lorenz number showing simple deviation from $L_0$[23], $0.1L_0$[21], $1.1L_0$[26], $3.5L_0$[19], $7L_0$[22], and $22L_0$[24]. The exception is $L_{xy} \sim 10^5 L_0$ observed in single crystalline Li$_{0.9}$Mo$_6$O$_{17}$ conductor, where charge carriers are spatially confined to 1D consistent with Tomonaga-Luttinger liquid (TLL); the ratio of the thermal ($\kappa_{xy}$) and electrical ($\sigma_{xy}$) Hall conductivities $\kappa_{xy}/\sigma_{xy}T$, in the form of Hall Lorenz number $L_{xy}$, instead of simple $L$ was measured[25].

In addition to large WFL, these NPs show metallic to semiconducting and finally to a completely insulating behavior with an intermediate systematic crossover from n-type to p-type conduction with the decrease in crystallite size. They show ultralow thermal conductivity of 0.52 ± 0.05 Wm$^{-1}$K$^{-1}$ (is about $\frac{1}{175}$th of its bulk value) at 300 K, which is smaller than those of many of the well-known TE materials[27]. The colossal Seebeck coefficient ($S$) of 1.87 ± 0.07 mVK$^{-1}$ is observed that is larger than those of ionic and solid-state thermoelectric supercapacitors[28,29]. These interesting nanoscale transport properties are mainly attributed to the extensive defects present due to the insulating TOP coated on the surface of the metallic core NPs[16,17], size effects and their collective response[2,5].

**EXPERIMENTAL METHODS**

**Synthesis of Ni Nanoparticles.** Nickel acetylacetonate, Ni(acac)$_2$ (95%), trioctylphosphine (TOP, 90%) and oleylamine (OA, 70%), purchased from Sigma Aldrich, were used as received. Typically, 3 g of Ni(acac)$_2$ and 10 ml of OA were mixed in three-neck round bottom flask and heated at 210 °C for 2 hours under nitrogen atmosphere. The reaction product was cooled down to room temperature and centrifuged after the addition of n-hexane and ethanol to extract the NPs. This was done three times to remove excess OA or acetate and then the particles were dried at 60 °C for characterizations. This sample was coded as Ni1. Then, second sample coded as Ni2 was prepared. For this, 0.25 ml of preheated TOP at 200 °C was added in a solution of 3 g of Ni(acac)$_2$ in 10 ml OA. The later solution was already degassed at 120°C for 30 mins with remaining reaction conditions the same as in Ni1. Similarly, samples coded as Ni3, Ni4, Ni5, Ni6, and Ni7 were prepared in 1 ml, 3 ml, 5 ml, 7 ml, and 10 ml TOP, respectively instead of 0.25 ml TOP used in Ni2; it may be noted herefrom that the amount of TOP in the samples preparation is highly influential to the novel properties of the compacted naoscale Ni. Details are given in table 1.

**Characterizations.** X-ray diffraction (XRD) measurements of Ni NPs were performed on powder samples using Bruker D8 Advance X-ray diffractometer with Cu Kα radiation (0.154 nm) for laboratory source and beamline BL-18, KEK, Japan for synchrotron radiation source. Transmission electron microscopy (TEM) and high-resolution TEM (HRTEM) measurements were performed using TECHNAI-20-G$^2$ on NPs dispersed over carbon-coated TEM grids by drop-casting the well-sonicated NPs. Field emission scanning electron microscopy (FESEM) measurements were carried out on compacted pellets using FEI Nova nanosem450. Resistance measurements were performed using four-point probes for Ni1 to Ni5 and two-point probes for Ni6 and Ni7[30]. Seebeck coefficient measurements using differential direct current setup in the temperature range of 5–300 K in a specially designed commercially available Dewar[30]. Thermal conductivity was measured using a dc pulse laser technique in the temperature range of 10 - 300 K[31]. Typical uncertainty in the measurements of these physical parameters were less than 3, 4 and 10%, respectively. These transport measurements were done on compacted pellets made from cold-pressed powder at 1.9 GPa and room temperature. The mass density of these pellets found to be greater than 77 % of the bulk value (8.46 g/cm$^3$).

**RESULTS and DISCUSSION**

**Structural and morphological study.** Ni NPs used here namely Ni1 through Ni7 were synthesized using thermal decomposition method as presented elsewhere[8] and it is briefly described in methods section; some of the details are in table 1. They show face-centered cubic (fcc) crystal structure of bulk Ni, without any impurity peak for Ni1 - Ni5. This has been confirmed from the Rietveld analysis (Fig. S1) of their laboratory X-ray diffraction (XRD) and synchrotron radiation beamline XRD patterns (Fig. 1a & inset). However, Ni6 and Ni7 were found to contain hexagonal closed packed structure along with fcc structure (table 1), in consistent with the earlier report[9]. XRD peaks are broadened with increase in TOP, manifesting decrease in crystallite size that is in line with the earlier report[8]. The average crystallite sizes as evaluated from Scherrer formula were 23.1 ± 0.3, 15.3 ± 0.2, 10.9 ± 0.4, 8.4 ± 0.2, 6.8 ± 0.3, 3.2 ± 0.2 and 1.3 ± 0.3 nm for Ni1, Ni2, Ni3, Ni4, Ni5, Ni6 and Ni7 (table 1); we use the terminology 'crystallite size' here to mean size of the particle in general throughout the paper as distinct from particle size (of say, TEM). TEM particle size may include non-crystalline outer region due to the interactions with surfactants. In TEM micrographs, there are random morphology and broad size distribution in Ni1 (Fig. 1b, c, inset), but narrower size distribution in Ni4 (Fig. 1d, e, inset) and Ni7 (Fig. 1f, g, inset) that exhibit reasonably monodispersed NPs with approximately similar morphology. The average TEM particle size is 70.5 ± 4.2, 10.9 ± 0.5 and 4.5 ± 0.3 nm with monodispersity of 51 %, 63 % and 88 % for Ni1, Ni4 and Ni7, respectively (table 1).

**Table 1** Sample preparation conditions, crystallite size, TEM size and lattice parameters of the samples.

| Sample | OA (ml) | TOP (ml) | Crystallite Size (nm) | TEM size (nm) | Lattice parameter (Å) |
|---|---|---|---|---|---|
| Ni1 | 10 | Nil | 23.1 ± 0.3 | 70.5 ± 4.2 | 3.5325(7) |
| Ni2 | 10 | 0.25 | 15.3 ± 0.2 | - | 3.5314(9) |
| Ni3 | 10 | 1 | 10.9 ± 0.4 | - | 3.530(2) |
| Ni4 | 10 | 3 | 8.4 ± 0.2 | 10.9 ± 0.5 | 3.528(2) |
| Ni5 | 10 | 5 | 6.8 ± 0.3 | - | 3.528(3) |
| Ni6 fcc hcp | 10 | 7 | 3.2 ± 0.2 - | - - | 3.528(6) a=b=2.650(1), c=4.360(9) |
| Ni7 fcc hcp | 10 | 10 | 1.3 ± 0.3 - | 4.5 ± 0.3 - | 3.523(7) a=b=2.621(1), c=4.322(1) |

As the quantity of TOP increases, the particle size decreases, and size distribution becomes narrower; the trend is consistent with crystallite size (table 1). Increase in broadening of XRD peaks with increase in quantity of TOP corroborates with the decrease in the degree of crystallinity with drop in particle size as obtained from selected area diffraction patterns (SAED) (Fig. 1c,e,g) and high-resolution TEM (HRTEM) of Ni1, Ni4 and Ni7 (Fig. 2a - c). HRTEM image of Ni1 shows the nonperiodic region or discontinuities between crystallites or core-shell including surfactant and grain boundaries and periodic regions of particle core (Fig.

2a). The inter-planner distance (d) is found to be ~ 0.204 nm corresponding to (111) plane (Fig. 2a).

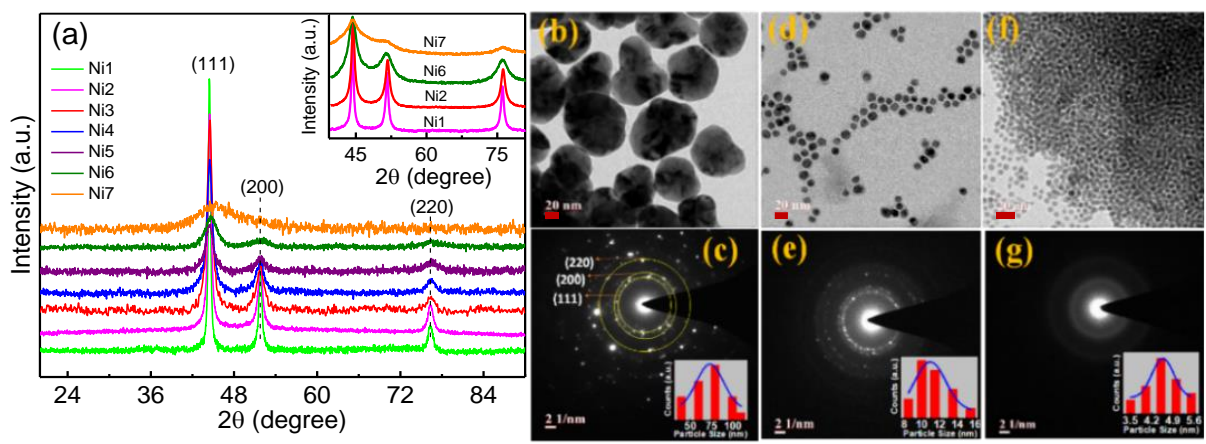

Fig. 1. (a) Laboratory source X-ray diffraction patterns of Ni1 through Ni7. Inset: synchrotron radiation x-ray diffraction patterns of Ni1, Ni2, Ni6 and Ni7. Transmission electron microscopy images (b, d, f) and selected area electron diffraction patterns (c, e, g) of Ni1, Ni4 and Ni7, respectively, scale bar of (b, d, f) is 20 nm; insets in c, e, g represent their size distributions.

The degree of crystallinity is suppressed in Ni4 relative to Ni1 that exhibits discloations or disturbed periodic region within the core (Fig. 2b). Particles become relatively smaller and less crystalline in Ni7 compared to Ni4 (Fig. 2c). Fig. 2d-f shows their respective FESEM images. Bigger particles in micron sizes with random morphology in Ni1 can be seen, while relatively smaller size particles are found in Ni4. Smaller round-shaped particles with size ranging from 60 nm to 100 nm are found in Ni7 (Fig. 2f), wherein size distribution is narrower as compared to Ni1 and Ni4 (Fig. 2d,e). This is consistent with TEM data (Fig. 1b, d, f).

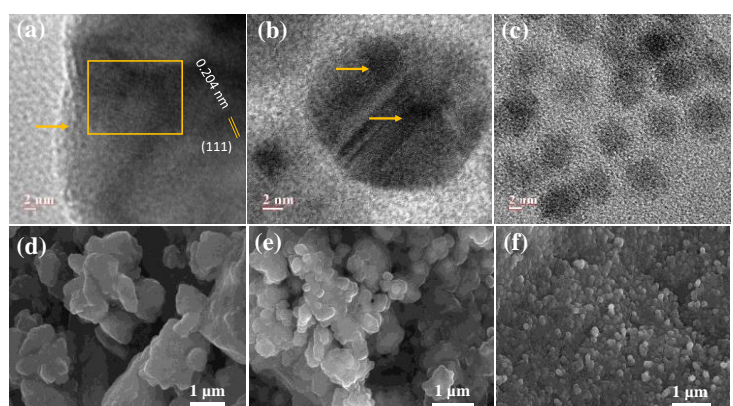

**Fig. 2.** (a,b,c), High resolution transmission electron microscopy and (d,e,f), field emission scanning electron microscopy images of Ni1, Ni4 and Ni7, respectively. Scale bar of a-c is 2 nm and that in d-f is 1 µm. Arrow and box in (a) show the nonperiodic region including surfactant and grain boundary and periodic region of particle core, respectively. Arrows in (b) shows the dislocations or disturbed periodic region within the particle core.

**Metal to insulator transition.** Fig. 3 shows the systematic evolution of Metal to insulator transition (MIT) in the electrical resistivity $\rho$ in Ni1 through Ni7 as the crystallite size decreases. Ni1 and Ni2 show clear metallic behavior in 5 – 300 K (Fig. 3a). Ni3, Ni4 and Ni5 show metallic behavior down to near 70 K, 100 K and 170 K, respectively (Figs. 3a, b), which are well below the probable respective superparamagnetic blocking temperatures (say, 330 K, 260 K and 180 K, respectively)[14] of these compacted monodispersed NP samples. The temperature dependence of $\rho(T)$ in the metallic regimes above these $T$'s of Ni1, Ni2, Ni3, Ni4 and Ni5 is not linear but follows a superlinear or sublinear power law $\rho \sim T^\delta$ (or $\sim T^{1.21}$) with the exponent $\delta$ = 1.20, 1.24, 1.73, 1.33, and 0.56, respectively (Fig. S2 and table S1). They are deviated from the standard $T^2$ dependence of a Fermi-liquid[32] and the sublinear $\rho \sim T^{0.56}$ behavior of Ni5 exactly matches with the chain or a-axis $\rho$ (i.e., $\rho_a$) of $(TMTSF)_2PF_6$ from 100 to 300 K with $K_c$ = 0.22, where $K_c$ is the Luttinger-liquid exponent controlling the decay of all correlation functions[33]. Below these temperatures, $\rho$ becomes semiconducting i.e., it increases with decrease in $T$. They therefore exhibit a temperature-driven MIT, which shifts to higher $T$ with decrease in crystallite size (table 1). The MIT completely disappears in Ni6 and Ni7 showing entirely semiconducting nature in the whole range of 10 to 300 K with enhanced feature to nearly insulating behavior in Ni7 (Fig. 3c). The evolution of this MIT may indicate that the electrical transport is via phonon-assisted tunneling or variable range hopping (VRH) rather than the long range hopping among the NPs through the grain boundaries (GBs) and surfactant matrix, wherein single electron charging becomes significant. It may result in Coulomb blockade behavior and localization of carriers on individual particles[5,34]. Similar MITs having Efros-Shklovskii VRH (ES-VRH) conduction have been reported in Al-Ge Films[35], granular Al[36,37], NbN film[38], Ni-SiO$_2$ films[39], Be films[40], Au nanoparticle multilayers[5], and Au nanocrystal arrays[3,4,34,6,41]. However, the magnitude of $\rho$ observed here is unprecedented. It shows a logarithmic rise to sixteen and to ten orders of magnitudes in Ni7 compared to Ni1 at 10 K and 300 K, respectively (Fig. 3d); $\rho$ at 10 K increases by 17 orders of magnitude in Ni7 relative to bulk Ni[18]. The residual resistivity ration (RRR) defined as the ratio of $\rho$ at 300 K, $\rho_{300K}$ and 10 K, $\rho_{10K}$, i.e. $\rho_{300K}/\rho_{10K}$ decreases with decrease in crystallite size (Fig. 3e), indicating enhanced disorder due to extensive defects as crystallite size reduces. Temperature coefficient of resistivity ($TCR = \frac{d\rho}{\rho dT}$) in the range of 10 K to 300 K is positive for Ni1 and Ni2 (Fig. 3f), but negative below 70 K, 100 K and 170 K for Ni3, Ni4 and Ni5, respectively due to the semiconducting nature, and simply semiconducting nature in the whole $T$ range in Ni6 and Ni7. The absolute value of $TCR$ systematically increases with decrease in crystallite size and attains a negative value of around -1.1 K$^{-1}$ at 10 K in Ni7 (Fig. 3g) that indicates onset of insulating phase below about 8 nm as depicted in Fig. 3h.

In order to get deeper insight into the semiconducting electrical transport mechanisms in different temperature regimes, ES-VRH, Mott-VRH and Arrhenius types of electrical transports viz., $\rho \sim \exp(T_{ES}/T)^{0.5}$, $\rho \sim \exp(T_M/T)^{0.25}$ and $\rho \sim \exp(E_a/k_BT)$, respectively were fitted in the semiconducting regimes of the $\rho$ of the compacted NP samples (Fig. S3), where $T_M = \frac{18}{\xi^3 N(E_F)k_B}$ and $T_{ES} = \frac{2.8e^2}{\varepsilon\xi}$ are the characteristic temperatures for the Mott-VRH and the ES-VRH conductions, respectively, with $E_a$ is the activation energy, $N(E_F)$ is the electronic density of states (DOS) at the Fermi level, $\xi$ as the localization length and $\varepsilon$ is the dielectric constant of the material. These fittings were essential since fitting of Arrhenius model of conduction alone was not satisfactory. The physical parameters thus obtained ($T_M$, $T_{ES}$ and $E_a$) from the fittings and calculated physical parameters of Coulomb gap energy ($\Delta_{CG}$), localization length ($\xi$), hopping length ($R_{hop}$) and charging energy ($E_c$) are shown in table 2. There is a crossover from the ES-VRH type of conduction to the Mott-VRH type of conduction and finally to the Arrhenius behavior with the increase in temperature. The crossover point shifts at higher $T$ with decrease in crystallite size (Fig. 3h & table 1).

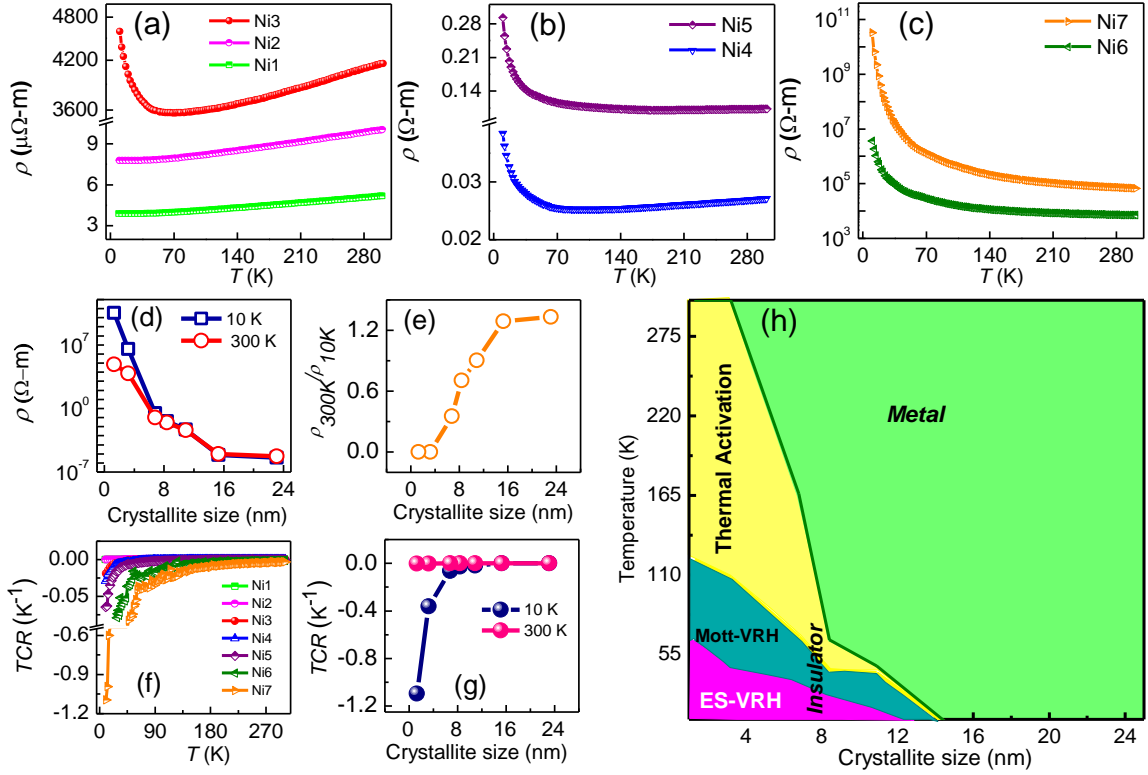

Fig. 3. Electrical resistivity $\rho$ of (a) Ni1, Ni2 & Ni3, (b) Ni4 & Ni5 and (c) Ni6 & Ni7. (d) Resistivity at 300 K ($\rho_{300K}$) and at 10 K ($\rho_{10K}$), and (e) $\rho_{300K}/\rho_{10K}$ as a function of crystallite size. (f) Temperature coefficient of resistivity (*TCR*) for Ni1 to Ni7, (g) *TCR* at 10 K and 300 K versus crystallite size. (h) Schematic illustration of metal to insulator phase transition versus the crystallite size.

The temperature regime, and the crossover temperature $T_C$ from the Mott-VRH conduction to the ES-VRH conduction[42] can be determined from either $T_C = 16\frac{T_{ES}^2}{T_M}$ as calculated $T_{C,cal}$ using the values of the Mott temperature $T_M$ and Efros-Shklovskii temperature $T_{ES}$ from table 2, or as experimental $T_{C,exp}$ from the point where the ES-VRH conduction deviates from the Mott-VRH conduction using the fitting of $\rho$ for Ni4 and Ni5 (table 3 & Fig. S3a-f). The $T_{C,exp}$ increases with decrease in crystallite size (table 1), in excellent agreement with the insulating behavior of smaller size NPs[42]. Notably, while $T_{C,cal}$ is close to the experimental $T_{C,exp}$ for Ni4 and Ni5, those for Ni6 and Ni7 are significantly different. Although $T_C$ is higher in Ni6 and Ni7 corresponding to coexistence of both the ES-VRH and the Mott-VRH models of electrical conductions, it is still within the range of investigated temperature i.e. 5 – 300 K. This behavior is similar to that of doped semiconductor[42] but is in contrast to the $T_C$ of ~1400 K in Au NPs[5]. Remarkably, the crossover of the model of electrical transport from the ES-VRH to the Mott-VRH conduction in Ni4 - Ni7 manifests opening of the Coulomb gap $\Delta_{CG} \approx \frac{e^3 N(E_F)^{\frac{1}{2}}}{\varepsilon^{\frac{3}{2}}} \approx k_B \left(\frac{T_{ES}^3}{T_M}\right)^{\frac{1}{2}}$ at the Fermi level, and $\Delta_{CG}$ thus estimated using these obtained parameters increases as crystallite size decreases[42] (table 2). We have calculated the localization length $\xi = \frac{e^2}{4\pi\varepsilon\varepsilon_0 T_{ES}}$ and the hopping length[5] $R_{hop} = \sqrt{\frac{e^2 \xi}{4 k_B T \pi \varepsilon \varepsilon_0}}$ at 10 K using the refractive index $\mu = 1.468$ for TOP obtained from the relation of the dielectric constant $\varepsilon = \mu^2 \sim 2.2$.

**Table 2.** ES-VRH temperature ($T_{ES}$), Mott-VRH temperature ($T_M$), calculated Coulomb gap energy ($\Delta_{CG}$), localization length ($\xi$), hopping length ($R_{hop}$), activation energy ($E_a$) and charging energy ($E_c$) of the samples in the semiconducting regimes.

| Sample (Crystallite size, nm) | $T_{ES}$ (K) | $T_M$ (K) | $\Delta_{CG}$ (meV) | $\xi$ (nm) | $R_{hop}$ (nm) | $E_a$ (meV) | $E_c$ (meV) |
|---|---|---|---|---|---|---|---|
| Ni3 (10.9) | 1.5 ± 0.1 | 2.3 ± 0.2 | - | 5400 | 4510 | 0.34 ± 0.05 | - |
| Ni4 (8.4) | 2.8 ± 0.2 | 6.0 ± 0.2 | 0.16 | 2700 | 3190 | 0.62 ± 0.06 | 118 |
| Ni5 (6.8) | 14.4 ± 0.3 | 110.2 ± 0.8 | 0.44 | 527 | 1410 | 1.4 ± 0.1 | - |
| Ni6 (3.2) | 671.1 ± 1.3 | 91709 ± 300 | 4.92 | 11.3 | 201 | 10.7 ± 0.4 | - |
| Ni7 (1.3) | 2745.0 ± 10.3 | 1151205 ± 2000 | 11.53 | 2.7 | 101 | 26.1 ± 0.8 | 297 |

**Table 3**. Calculated and fitted values of crossover temperature ($T_C$) from the ES-VRH to the Mott-VRH of conduction, ratios of hopping length ($R_{ES}$) to localization length ($\xi$) in ES-VRH ($\frac{R_{ES}}{\xi}$) and hopping length ($R_M$) to localization length ($\xi$) in Mott-VRH ($\frac{R_M}{\xi}$) conduction and fitted values of the exponent $x$ and calculated values of exponent $\gamma$ of the samples in the semiconducting regimes.

| Sample (Crystallite size, nm) | $T_{C,exp}$ (K) | $T_{C,cal}$ (K) | $\frac{R_{ES}}{\xi}$ | $\frac{R_M}{\xi}$ | Fitted $x$ | Calculated $\gamma$ |
|---|---|---|---|---|---|---|
| Ni3 (10.9) | - | - | 0.306× $T^{-1/2}$ | 0.46× $T^{-1/4}$ | - | - |
| Ni4 (8.4) | 27 | 21 | 0.418× $T^{-1/2}$ | 0.58× $T^{-1/4}$ | 0.53 ± 0.04 | 2.43 ± 0.04 |
| Ni5 (6.8) | 34 | 30 | 0.948× $T^{-1/2}$ | 1.21× $T^{-1/4}$ | 0.54 ± 0.03 | 2.52 ± 0.03 |
| Ni6 (3.2) | 47 | 82 | 6.470 × $T^{-1/2}$ | 6.52× $T^{-1/4}$ | 0.52 ± 0.05 | 2.25 ± 0.05 |
| Ni7 (1.3) | 61 | 104 | 13.09× $T^{-1/2}$ | 12.28×$T^{-1/4}$ | 0.50 ± 0.02 | 2.00 ± 0.02 |

The value of $\xi \sim 2.7$ nm obtained (Table 2) using its average TEM particle size of 4.5 nm (Table 1) for Ni7, implies localization of charge carriers (see Fig. 3c), while its $R_{hop}$ at 10 K is around 37 times larger than $\xi$ and can cover up to 22 NPs of 4.5 nm (table 2). This is in excellent agreement with that of Au NPs[4,29], and indicates that carriers are transported through cotunneling at low temperature; these neutral excitations do not transport charge[20]. Similar transport is indicative from $R_{hop} > \xi$ in Ni4 – Ni6 (table 2); $\varepsilon$ =2.2 was used for other samples also to calculate their $\xi$ and $R_{hop}$ at 10 K since OA has $\mu \sim 1.460$, comparable to that of TOP. However, hopping conduction in higher $T$ is expected for Ni3 since $\xi > R_{hop}$. This is consistent with small change in $\rho$ in the semiconducting regime (at 10 K) relative to metallic regime (300 K). Overall, while there are decrease in both $\xi$ and $R_{hop}$ with decrease in crystallite size, there is increase in the $T_{ES}$, $T_M$ and $\Delta_{CG}$ parameters (table 2). Further evidences on cotunneling transport are $\frac{R_{ES}}{\xi} > 1$ in Ni7 and $\frac{R_{ES}}{\xi} \sim 1$ in Ni6, and $\frac{R_M}{\xi} > 1$ in Ni6 and Ni7 (table 3); they are determined from the ratio of hopping length ($R_{ES}$ or $R_M$) to $\xi$ i.e. $\frac{R_M}{\xi} = \frac{3}{8}\left(\frac{T_M}{T}\right)^{\frac{1}{4}}$ for Mott-VRH ($R_M$) or $\frac{R_{ES}}{\xi} = \frac{1}{4}\left(\frac{T_{ES}}{T}\right)^{\frac{1}{2}}$ for ES-VRH ($R_{ES}$) conduction[42] using $T$ as $T_{c,exp}$.

The approximate Coulomb charging energy $E_C = \frac{e^2}{4\pi\varepsilon\varepsilon_0 r}$ for single NP with radius $r$ was also calculated using TEM particle sizes for Ni4 and Ni7 as $E_c \sim 297$ eV and 118 eV, respectively (table 1). $E_c \sim 297$ meV for Ni7 is almost 12 times that of thermal energy $E_a \sim 26.1$ eV at 300 K, and suggests strong Coulomb blockade behavior implying that electrons inside the crystallite will create a strong Coulomb repulsion preventing other electrons to flow when the crystallite is small enough. This is more pronounced (190 times) in Ni4; $E_C$ may be overestimated but cannot be lower than $E_a \sim 26$ meV. Now, we use the prediction of the ES-VRH conduction on the DOS at Fermi level[42,43] that varies as[2] $N(E) = N_0|E - E_F|^\gamma$; more details on this expression are discussed in SI. The critical exponent ($\gamma$) for variation in DOS at the Fermi energy using this expression lies approximately between 2 to 2.5 (Fig. S4). It decreases with decrease in crystallite size (table 3) that is in excellent agreement with earlier reports[2,43]. The value of $\gamma = 2$ for Ni7 well-matches with its theoretical value $\gamma = 2$ for 3D. The slightly larger other values are consistent with increase in Mott temperature $T_M$ and Efros-Shklovskii temperature $T_{ES}$ with decrease in crystallite size. Therefore, while cotunneling transport at low temperature gives rise to ES-VRH conduction, increase in $T$ gives the Mott-VRH conduction and finally to the thermally activated type of conduction (Fig. 3h). This clearly reveals that there is Coulomb gap, an essential ingredient of large violation of WFL[3,25], in the ES-VRH conduction where DOS varies nearly in quadratic form i.e. $\gamma = 2$ in the smallest particle size sample, which is consistent with earlier report on the array of metal nanocrystals[25]. These ES-VRH, and Mott-VRH types of electrical conductions in addition to Arrhenius type at various metallic T-regimes, not at too low temperatures, appear distinct from that of the nonmagnetic granular metal[20]. This may reflect the intricate properties of these superparamagnetic granular NPs[8,9,10–12,13,14]. To correlate the electrical transport properties, their thermal conductivity $\kappa$ was investigated.

**Ultralow thermal conductivity ($\kappa$).** The $\kappa$ varies monotonically with $T$ associated with a broad hump-like feature at around 200 K in Ni1 (Fig. 4a), which is attributed to enhanced scattering of charge carriers with defects, consistent with Fig. 3e. It drops quite significantly with decrease in crystallite size, consistent with single Ni nanowire[26]. However, such trend has been changed in Ni3 with a peak near 32 K with a broad hump-like feature still preserving at around 200 K. This peak near 32 K evolves gradually more distinctly in smaller NP size samples, and it is the sharpest in Ni7 while hump-like feature slowly disappears turning into a straight line as the crystallite size decreases (Fig. 4a, inset for clarity). The peak near 32 K is tentatively attributed to monodispersity and nanolattice formation of the nanoparticles[8], which is in line with its small value unlike that in metals where its values are relatively very large.

The value of $\kappa$ at 300 K falls fast nearly linearly but that at 10 K increases slowly and then drops again as the crystallite size decreases (Fig. 4b). Their trends are distinct from those of $\rho$ (Fig. 3d), but consistent with earlier theoretical report on Ni NPs[44]. There are thus tendencies of evolution of decreasing $\kappa$ values, changing shape and slope as the crystallite size decreases. While the lowest value of $\kappa$ is ~ 0.13 ± 0.01 W/m-K at 10 K, it is ~ 0.52 ± 0.05 W/m-K at 300 K in Ni7. The latter value is around 1/175 of ~ 91.0 W/m-K of Ni bulk[44], showing its ultralow value. This is significantly smaller than the well-known TE materials (Fig. S5a) such as nanocrystalline $Si_{80}Ge_{20}$ and $Bi_2Te_3$, PbS, $TiS_2$ and $(Cu,Fe)_5S_4$ [ref.[45] and references therein]. In overall, the $T$-dependence of $\kappa$ is intriguing. This is realized from the expression $\kappa \sim T^y$ that shows the values of the exponent $y$ ranging from 0.7 to 1.3 depending on crystallite size at low temperature (Fig. S5b). This is consistent with anatase $TiO_2$[46]. In the metallic higher $T$ regimes, $\kappa$ exhibits exponential dependence of $T$ as $\kappa \sim e^{-\beta T}$ (or $\sim e^{-0.0058T}$), where the decay constant $\beta$ = 0.0048, 0.0096, 0.0032 and 0.0056 for Ni1, Ni3, Ni4 and Ni5, respectively (table S2 and Fig. S5c). These are attributed to the multifaceted properties of these usually superparamagnetic granular Ni NPs[8,9,10–12,13,14] in contrast to the theoretical phonon-like $T^3$-dependence at low $T$ for granular metal[20]. The sharp drop in $\kappa$ at low temperature is likely to be due to the strong grain boundary scattering of charge carriers and localization effects associated with a decrease in $\xi$ with decrease in size and scattering of carriers with defects including point defects, dislocations and GBs (Figs. 1b-g & 2a-c). They may play important roles in the electrical transport in Ni3 to Ni7 with the crystallite size smaller than the mean free path of the electrons (~ 14 nm) in bulk Ni[47].

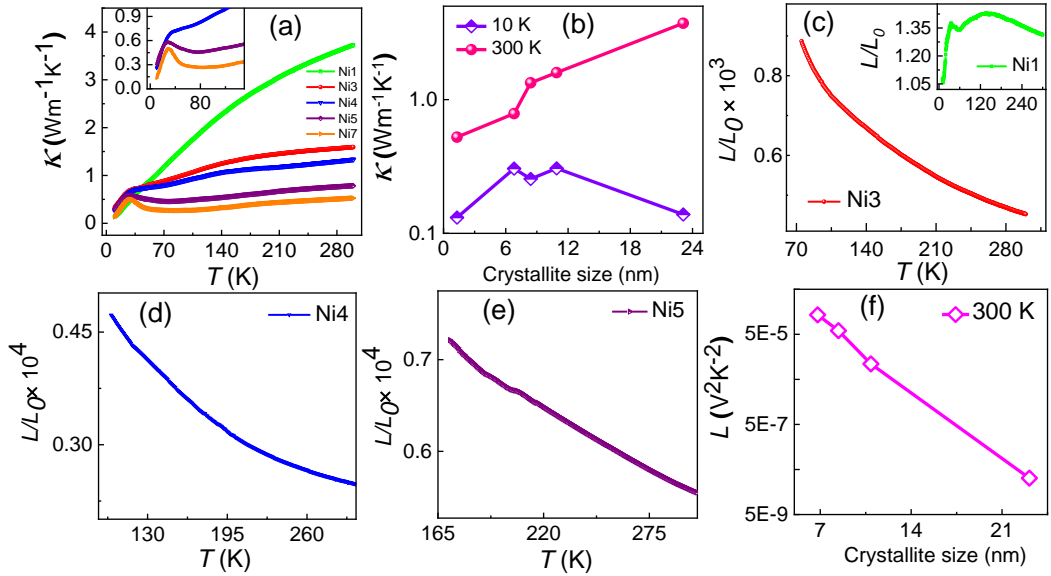

Fig. 4. (a) Thermal conductivity for Ni1, Ni3, Ni4, Ni5 and Ni7. (b) Thermal conductivity at 10 K and 300 K as a function of crystallite size. Ratio of effective Lorenz number for the nanoparticle ($L$) to Sommerfeld value ($L_0$) i.e., $L/L_0$ in metallic regime of (c) Ni3; inset shows the $L/L_0$ of Ni1, (d) Ni4 & (e) Ni5. (f) L versus crystallite size at 300 K.

**Lorenz number (*L*).** We have calculated Lorenz number of these NPs using the WFL, wherein $L=\frac{\kappa_e}{\sigma T}$ in the metallic regimes only, in consistent with literature[25]. However, there exist some reports on metallic systems in which $\kappa$ is used directly in place of $\kappa_e$ to calculate *L* by simply neglecting the phonon thermal conductivity $\kappa_{ph}$ [23,26,48]. This may lead to additional enhancement in *L* since $\kappa = \kappa_e + \kappa_{ph}$. Because accurate determination of *L* for metals with short mean-free-paths or metallic nanostructures is difficult since both the $\kappa_e$ and $\kappa_{ph}$ can be comparable in magnitude[13,49]. Moreover, theoretical calculation and experimental determination of $\kappa_{ph}$ has not therefore been possible and hence has not been reported so far for nanoparticles or bulk nanostructured materials. We have attempted here to estimate the $\kappa_{ph}$ by using some standard approaches. Usually, $\kappa_{ph}$ is given by $\kappa_{ph} = \frac{1}{3}C_v v l$, where $C_v$, $v$ and $l$ are specific heat, sound velocity and mean free path of phonons. For NPs, $v$ can be obtained by using the relation $v = \frac{\theta_D v_B}{\theta_{DB}}$, where $\theta_D$ is Debye temperature for NPs, and $\theta_{DB}$ and $v_B$ are Debye temperature and sound velocity, respectively for bulk material[50]. Even though both these parameters depend on the particle size and hence velocity, we have taken $\theta_D \sim 269$ K and $C_v = 0.27$ J/g.K at 300 K[13] for ~ 6 nm to 10.1 nm of such Ni NPs[13]. The values of $v_B$ and $\theta_{DB}$ for bulk Ni are 5630 m/s and 453 K, respectively[51]. From the above relation and reported parameters, we have calculated $v \sim 3343$ m/s for these NPs. Even though the mean free path *l* is ~ 1.2 nm at 300 K for bulk Ni[51], it would be less than this value in such disordered NPs due to phonon-boundary and phonon-impurity scattering, specifically point defects and dislocations. The *l* may finally tend to the lattice parameter ~ 0.3528 nm. It is to be noted that both *v* and *l* have temperature dependence and their exact calculation for the NPs is thus not possible. With the above parameters, $\kappa_{ph} \sim 0.72$ Wm$^{-1}$K$^{-1}$ at 300 K for Ni3. The $\kappa_{ph}$ found is about 50 % of total $\kappa$ in these Ni NPs, which is relatively larger compared to some of the earlier reports[26,51] but well-comparable with 35-40% at 300 K in others like metallic VN$_x$ epitaxial layers[49]. Based on the 50 % phonon contribution to the thermal conductivity in metallic nanostructures, we have considered $\kappa_e = \kappa_{ph}$ for metallic regimes of these Ni NPs to calculate the effective *L*.

Figs 4c-e depicts the normalized Lorenz ratio *L/L₀* thus determined as a function of temperature (*T*) for Ni3, Ni4 and Ni5; Figs 4c, inset shows this for Ni1. Remarkably, while the ratio *L/L₀* ranges around 1 - 1.25 in the metallic Ni1, it increases to large values of ~ 0.4 – 7 × 10³ for the MIT-induced Ni3, Ni4 and Ni5 in their metallic regimes. Fig. 4f shows *L* dependence on crystallite size at 300 K. It clearly shows a significant deviation from WFL or FL behavior. This WFL violation is marginal in metallic Ni1 but very large in the metallic regimes of Ni3-Ni5. The former looks consistent with the theoretically large $g \gg 1$ but probably not those of Ni3-Ni5 even for $g \ll 1$ with just a qualitative serious WFL violation[20] since the large Lorenz numbers observed here are unprecedented compared with those reported so far with a maximum of *L* as 22 $L_0$[24], with the exception of *Hall L* ~ 10⁵ $L_0$ observed in quasi-1D Li$_{0.9}$Mo$_6$O$_{17}$ conductor[25]. The violation magnitude for metallic Ni1 is comparable to the experimental single Ni nanowire[26] and to the theoretical report on nanowire of Pt and Ag[19], and Au nanofilms[23,48]. The large violation of the WFL observed here indicates a dramatic change in the excitation spectrum accompanied with the opening of a gap[3] and the conducting ground state of these granular Ni NPs is perhaps simply not an FL[32,24,25,52,53]. The $T$-dependence of $L(T)/L_0$ in the metallic regimes of $\rho$ follows varied power laws as $L(T)/L_0 \sim T^{1.16}$ for Ni1, $\sim T^{0.05}$ for Ni3, $\sim T^{0.08}$ for Ni4, and $\sim T^{1.66}$ for Ni5 (Fig. S6 and table S3) indicating the complex nature of the decoupled transports of charge and spin excitations as crystallite size changes[32,24,25,52,53]. They are associated with the ten orders of magnitude of enhancements in $\rho$ (Fig. 3d) in contrast to just less than an order of magnitude increase in $\kappa$ at 300 (Fig. 4b) in the enhanced disordered Ni7 (Fig. 3e) compared to Ni1. The ES-VRH, Mott-VRH and Arrhenius types of electrical conductions in the various semiconducting regimes as $\rho$ approaches the MIT are associated with the opening of an energy gap, cotunneling, Coulomb blockade and transport energy degradation of the charge carriers. The $\rho \sim T^{1.21}$ and $\kappa \sim e^{-0.0058T}$ behaviors in the metallic $T$ regimes of these well-known superparamagnetic Ni NPs[14] show distinction from those of the theoretical $\kappa \sim T^3$ domination by cotunneling of low-energy particle-hole pairs on the nonmagnetic granular metal at low temperatures[20]; see more details in the discussion. The cotunneling of electron-hole pairs can transport heat, but not charge from one grain to another since they are neutral excitations in line with the highly enhanced $\rho$ (Fig. 3) and $E_c$ (Table 2). Thus, these conduction mechanisms can enhance the thermal conductivity significantly compared to electrical conductivity that leads to increase in $L$[20,22]. Therefore, there is conservation of grain charge, and non-conserving energy perhaps decouple the charge and heat conductivities to violate the WFL significantly[3,20]. They are mainly attributed to the extensive defects present as evident from RRR drop (Fig. 3e) due to the prevailing metal-organic interfaces, volume fractions of nanocrystal cores, surface ligand, other defects including GBs and point defects. They may be due to their intricate properties[8,9,10–12,13,14] and can significantly modify the heat and charge transport[54,55] as crystallite size drops (also see Ref.[21]). Another possibility may be the bipolar diffusion[22] as evident in Seebeck coefficient.

**Colossal Seebeck coefficient ($S$).** Fig. 5a-c shows the $S$ for Ni1 through Ni7 and annealed bulk polycrystalline Ni ingot (having purity 99.99%) as reference. The $S$ of bulk Ni exhibits a dip-like feature about 45 K due to phonon-drag minimum (PDM)[20], a broad hump near 120 K and approximately linear increment with $T$ due to charge diffusion (Fig. 5a)[12]. The PDM is systematically suppressed with the broad hump-like feature shifting at $\sim$ 140 K in Ni1 and Ni2 compared to Ni bulk and then, $S$ crosses over from negative to positive sign near 207 K and 275 K for Ni1 and Ni2, respectively, as $T$ increases (Fig. 5a)[20]. They are assigned to the change in overall relaxation rate of charge carriers due to their enhanced interactions with various defects as crystallite size drops[20,56].

These features are considerably transformed with further decrease in crystallite size with just a feeble phonon drag effect in each of Ni3 and Ni4 (Fig. 5b), and PDM has been completely disappeared in Ni5 to Ni7 (Fig. 5c). Consequently, large positive values are observed in the whole of $T$ range for Ni3 and Ni4, and colossal positive values for Ni5 to Ni7. They show evolution of electrical conduction from $n$-type to $p$-type and hence might indicate the possible bipolar diffusion in these NPs as the crystallite size drops. Fig. 5d illustrates the trends in $S$ of the samples at 10 K and 300 K that show approximately $10^4$ and 473 times increase of it, respectively, in Ni7 compared to that of Ni1; this value could be much larger when it is extrapolated to or at 10 K but considered just at 30 K for Ni7 (Fig. 5c). This enhancement in $S$ is not that large as that in $\rho$ where it is 17 orders of magnitude (Fig. 3d). The maximum values of $S$ are 68.6 ± 2.8 and 1.87 ± 0.07 mVK$^{-1}$ at 30 K and 300 K, respectively for Ni7. They are very large compared to other materials (Fig. S7a) including CoSbS[57], Cu$_2$Se[58], ionic and solid-state thermoelectric supercapacitors[28,29]. Notably, $S$ for Ni7 could not be measured below 30 K due to larger resistance than the input impedance ($10^{10}$ Ω) of nanovoltmeter used to measure $S$.

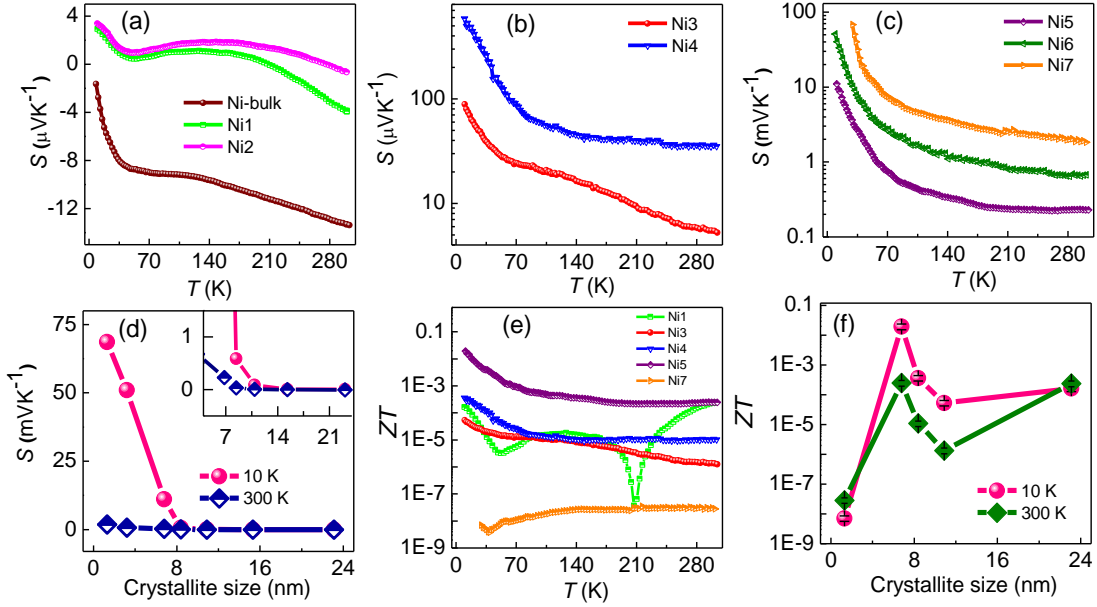

Fig. 5. Seebeck coefficient (a) Bulk Ni, Ni1 & Ni2, (b) Ni3 & Ni4 and (c) Ni5, Ni6 & Ni7. (d) Seebeck coefficient at 10 K and 300 K as a function of crystallite size; inset: expanded view at smaller scale. Thermoelectric figure of merit $ZT$ as a function of (e) temperature and (f) crystallite size at 30 K & 300 K of Ni1, Ni3, Ni4, Ni5 and Ni7.

The $S$ for metals or degenerate semiconductors is associated with derivative of energy-dependent carrier concentration $n$ ($E$) and mobility $\mu(E)$, which may be positive or negative depending on their electronic structure[59,60]. We believe that crossover from negative to positive sign is mainly appeared due to modification in electronic structure with reduction in particle size, which is consistent with earlier report[56]. Notably, the steep increase in $S$ at low $T$ can be correlated with steady increase in $\rho$ that is attributed to opening of the Coulomb gap ($\Delta_{CG}$) at the Fermi level at low $T$. The drop in carrier concentration $n$ is evident from sharp increase in $\rho$ (Fig. 3), consistent with the possible bipolar transport and hence, increase in $\Delta_{CG}$ as the crystallite size drops (table 2). This enables the increase in S since it is inversely related to $n$[59,60]. The analysis of $S$ reveals the $p$-type conduction and $\rho$ insulating behavior with decrease in crystallite size. This is consistent with their electronic structure obtained from x-ray absorption spectroscopy similar to $p$-type material NiO as size decreases[9].

Thus, considering these nanoparticles as $p$-type semiconductors, their thermopower can be given as[61,62]

$S = \frac{k_B}{e}\left(\frac{E_F - E_T}{k_B T} + \alpha\right),$ (1)

where $E_F$, $E_T$ and $\alpha$ are Fermi energy level, transport level, and heat of transport constant, respectively. Notably, α is related to the width of the distribution of DOS. This is related the steepness of its rise from $E_T$ to the maximum of distribution of DOS[62]. If the particle size is bigger and has broad size distribution, then their DOS profile will be broader and $S$ depends on where the $E_T$ exists (Fig. S7b). Generally, the $E_T$ in a nanocrystalline metal may exist between the peak and edge of DOS depending on the monodispersity of the NPs that determines the distribution of DOS. We have obtained the position of $E_T$ with respect to the $E_F$, i.e., $E_F$-$E_T$ from the fitting of $S$ versus $1/T$ for Ni3 to Ni7 (eq. 1)[62]. The transport energy level differences from the Fermi level ($E_F$-$E_T$), obtained from the slopes of the linear fits of $S$ versus $1000/T$ (Fig. S7c-e), are 3.0 ± 0.1 meV, 3.5 ± 0.2 meV, 15 ± 1 meV, 153 ± 5 meV and 396 ± 14 meV for Ni3, Ni4, Ni5, Ni6 and Ni7, respectively. They indicate fast fall in the transport energy with decrease in particle size[62], enhancing inelastic transport and hence deviating from the WFL. This therefore further supports the increase in $L$ with increase in ($E_F$-$E_T$) i.e., $E_T$ moves away from the Fermi energy with decrease in crystallite size. This is consistent with earlier computational study on violation of WFL[19]. The large positive values of $E_F$-$E_T$ support the large positive sign of $S$ in Ni5 - Ni7. This is in line with their insulating behavior of smaller particle size samples from Ni1 to Ni7. This seemingly leads to the sharp modulation in DOS near Fermi level and to the remarkable increase in $S$[62].

We have calculated the thermoelectric figure of merit ($ZT$), defines as $ZT = \frac{S^2 \sigma T}{\kappa}$ and power factor ($S^2\sigma$). $ZT$ (fig. 5e, f) and $S^2\sigma$ (fig. S8a, b) as function of temperature and crystallite size show similar trends. $ZT$ shows a very sharp dip near 207 K in Ni1 alone that relates to the crossover of $S$ from negative to positive values. It slightly rises below 50 K. For Ni7, $ZT$ saturates slightly, and then drops in the value at around 35 K and is the smallest amongst all the samples. The $ZT$ at 10 K and 300 K increases as particle size decreases from Ni1 to Ni3, then increases in Ni4 and Ni5 and smallest in Ni7 (Fig. 5f). The maximum values of $S^2\sigma$ and $ZT$ are found to be 450 ± 49 μW/m-K$^2$ and 0.020 ± 0.004, respectively in Ni5 at 10 K. In spite of Colossal $S$ and ultralow $\kappa$, these NPs show a low value of $ZT$ due to simultaneous remarkable enhancement in $\rho$ because of insulating nature of surfactant/s. However, the value of ZT is about two orders of magnitude larger than that of bulk Ni. This study, therefore, provides an intriguing platform to tune the charge and heat transport in nanoscale materials.

CONCLUSIONS

In conclusion, we have experimentally demonstrated a large violation of Wiedemann-Franz law, systematic metal to insulator transition, ultralow thermal conductivity, and large Seebeck coefficient in compacted colloidal Ni nanoparticles as size drops. They show some unconventional quasiparticle dynamics with change in conduction from *n*-type to *p*-type, and are attributed to cotunneling, Coulomb blockade, and drop in the transport energy due to the extensive defects present in these nanoparticles as the crystallite size reduces. These results therefore not only provide a window into the unconventional quasiparticle transport in such compacted nanoparticles where the transport of charge and heat is decoupled, but also may be useful for thermoelectrics and heat dissipation in nanocrystal array-based electronics.

ACKNOWLEDGMENTS
Authors gratefully acknowledge Er. Vipin Kumar, Council of Scientific & Industrial Research (CSIR)-Central Electronics Engineering Research Institute (CEERI), Pilani, Rajasthan, India for providing FESEM and TEM data through CIL, Harisingh Gour University, Sagar and MRC, MNIT, Jaipur, India, respectively. Authors are thankful to Dr. Mukul Gupta, UGC-DAE Consortium for Scientific Research, Indore, India for providing laboratory XRD data and Dr. Jaiveer Singh, IPS Academy, Indore, India for collecting synchrotron radiation XRD at KEK, Photon factory, Tsukuba, Japan.
REFERENCES

[1] I. S. Beloborodov, A. V. Lopatin, V. M. Vinokur, and K. B. Efetov, Rev. Mod. Phys. **79**, 469 (2007).

[2] T. Chen, B. Skinner, and B. I. Shklovskii, Phys. Rev. Lett. **109**, 126805 (2012).

[3] H. Moreira, Q. Yu, B. Nadal, B. Bresson, M. Rosticher, N. Lequeux, A. Zimmers, and H. Aubin, Phys. Rev. Lett. **107**, 176803 (2011).

[4] A. Zabet-Khosousi, P. E. Trudeau, Y. Suganuma, A. A. Dhirani, and B. Statt, Phys. Rev. Lett. **96**, 156403 (2006).

[5] T. B. Tran, I. S. Beloborodov, X. M. Lin, T. P. Bigioni, V. M. Vinokur, and H. M. Jaeger, Phys. Rev. Lett. **95**, 076806 (2005).

[6] T. B. Tran, I. S. Beloborodov, J. Hu, X. M. Lin, T. F. Rosenbaum, and H. M. Jaeger, Phys. Rev. B **78**, 075437 (2008).

[7] V. Sharma, D. Verma, G. S. Okram, R. J. Choudhary, D. Kumar, and U. Deshpande, J. Magn. Magn. Mater. **497**, 166000 (2020).

[8] J. Singh, N. Kaurav, N. P. Lalla, and G. S. Okram, J. Mater. Chem. C **2**, 8918 (2014).

[9] Tarachand, V. Sharma, J. Singh, C. Nayak, D. Bhattacharyya, N. Kaurav, S. N. Jha, and G.S. Okram, J. Phys. Chem. C **120**, 28354 (2016).

[10] G. S. Okram, A. Soni, and R. Rawat, Nanotechnology **19**, 185711 (2008).

[11] P. V. P. Madduri and S. N. Kaul, Phys. Rev. B **95**, 184402 (2017).

[12] A. Soni and G. S. Okram, Appl. Phys. Lett. **95**, 013101 (2009).

[13] J. Singh, Tarachand, S. S. Samatham, D. Venkateshwarlu, Netram Kaurav, V. Ganesan, and G. S. Okram, Appl. Phys. Lett. **111**, 201904 (2017).

[14] J. T. Batley, M. Nguyen, I. Kamboj, C. Korostynski, E. S. Aydil, and C. Leighton, Chem. Mater. **32**, 6494 (2020).

[15] G. Weidemann and R. Franz, Ann. Phys. **89**, 497 (1853).

[16] K.-S. Kim and C. Pépin, Phys. Rev. Lett. **102**, 156404 (2009).

[17] Y. Zhang, N. P. Ong, Z. A. Xu, K. Krishana, R. Gagnon, and L. Taillefer, Phys. Rev. Lett. **84**, 2219 (2000).

[18] A. Garg, D. Rasch, E. Shimshoni, and A. Rosch, Phys. Rev. Lett. **103**, 096402 (2009).

[19] A. Marius, Bürkle and Yoshihiro, Nano Lett. **18**, 7358 (2018).

[20] V. Tripathi and Y. L. Loh, Phys. Rev. Lett. **96**, 046805 (2006).

[21] S. Lee, K. Hippalgaonkar, F. Yang, J. Hong, X. Zhang, C. Dames, and S.A. Hartnoll, Science **355**, 371 (2017).

[22] Z. Luo, J. Tian, S. Huang, M. Srinivasan, J. Maassen, Y. P. Chen, and X. Xu, ACS Nano **12**, 1120 (2018).

[23] S. J. Mason, D. J. Wesenberg, A. Hojem, M. Manno, C. Leighton, and B. L. Zink, Phys Rev Mat. **4**, 065003 (2020).

[24] J. Crossno, J. K. Shi, K. Wang, X. Liu, A. Harzheim, A. Lucas, S. Sachdev, P. Kim, T. Taniguchi, K. Watanabe, T.A. Ohki, and K.C. Fong, Science **351**, 1058 (2016).

[25] W. Franz, N. Wakeham, A. F. Bangura, X. Xu, J. Mercure, M. Greenblatt, and N. E. Hussey, Nat. Commun. **2**, 396 (2011).

[26] M. N. Ou, T. J. Yang, S. R. Harutyunyan, Y. Y. Chen, C. D. Chen, and S. J. Lai, Appl. Phys. Lett. **92**, 063101 (2008).

[27] V. Sharma, G. S. Okram, D. Verma, N. P. Lalla, and Y. K. Kuo, ACS Appl. Mater. Interfaces

**12**, 54742 (2020).

[28] D. Zhao, H. Wang, Z. U. Khan, J. C. Chen, R. Gabrielsson, M. P. Jonsson, M. Berggren, and X. Crispin, Energy Environ. Sci. **9**, 1450 (2016).

[29] S. L. Kim, H. T. Lin, and C. Yu, Adv. Energy Mater. **6**, 1600546 (2016).

[30] A. Soni and G.S. and Okram, Rev. Sci. Instrum. **79**, 125103 (2008).

[31] Y. Kuo, B. Ramachandran, and C. Lue, Front. Chem. **2**, 106 (2014).

[32] R. W. Hill, C. Proust, L. Taillefer, P. Fournier, and R. L. Greene, Nature **71**, 711 (2001).

[33] M. Dressel, K. Petukhov, B. Salameh, P. Zornoza and T. Giamarchi, Phys. Rev. B **78**, 075104 (2005).

[34] T. Sugawara, M. Minamoto, M.M. Matsushita, P. Nickels, and S. Komiyama, Phys. Rev. B **77**, 235316 (2008).

[35] A. Gerber, A. Milner, G. Deutscher, M. Karpovsky, and A. Gladkikh, Phys. Rev. Lett. **78**, 4277 (1997).

[36] T. Chui, G. Deutscher, P. Lindenfeld and W. L. McLean, Phys. Rev. B **23**, 6172 (1981).

[37] N. Bachar, S. Lerer, B. Almog, G. Deutscher, S. Hacohen-Gourgy, B. Almog, and G. Deutscher, Phys. Rev. B **87**, 214512 (2013).

[38] R. W. Simon, B. J. Dalrymple, D. Van Vechten, W. W. Fuller and S. A. Wolf, Phys. Rev. B **36**, 1962 (1987).

[39] P. Sheng, B. Abeles and Y. Arie, Phys. Rev. Lett. **31**, 44 (1973).

[40] V. Y. U. Butko, J. F. DiTusa, and P. W. Adams, Phys. Rev. Lett. **84**, 1543 (2000).

[41] J. L. Dunford, Y. Suganuma, A. A. Dhirani, and B. Statt, Phys. Rev. B **72**, 075441 (2005).

[42] Y. Zhang, Peihua Dai, Miguel Levy, and M. P. Sarachik, Phys. Rev. Lett. **64**, 2687 (1990).

[43] R. Rosenbaum, Phys. Rev. B **44**, 3599 (1991).

[44] S. P. Yuan and P.X. Jiang, Int. J. Thermophys **27**, 581 (2006).

[45] P. Qiu, T. Zhang, Y. Qiu, X. Shi, and L. Chen, Energy Environ. Sci. **7**, 4000 (2014).

[46] X. Mettan, J. Jaćimović, O. S. Barišić, A. Pisoni, I. Batistić, E. Horváth, S. Brown, L. Rossi, P. Szirmai, B. Farkas, H. Berger, and L. Forró, Commun. Phys. **2**, 123 (2019).

[47] D. Gall, J. Appl. Phys. **119**, 085101 (2016).

[48] A. D. Avery, S. J. Mason, D. Bassett, D. Wesenberg, and B. L. Zink, Phys. Rev. B **92**, 214410 (2015).

[49] Q. Zheng, A. B. Mei, M. Tuteja, D. G. Sangiovanni, L. Hultman, J. E. Greene, and D. G. Cahill, Phys. Rev. Mater. **1**, 065002 (2017).

[50] M. Goyal, Pramana **91**, 87 (2018).

[51] S. Yuan and P. Jiang, Prog. Nat. Sci. **15**, 922 (2005).

[52] X. Han, B. Liu, and J. Hu, Phys. Rev. A **100**, 43604 (2019).

[53] C. L. Kane and M.P.A. Fisher, Phys. Rev. Lett. **76**, 3192 (1996).

[54] L. Wang, Z. Zhang, Y. Liu, B. Wang, L. Fang, J. Qiu, K. Zhang, and S. Wang, Nat. Commun. 9, 3817 (2018).

[55] W. Ong, S. M. Rupich, D. V Talapin, A. J. H. Mcgaughey, and J. A. Malen, Nat. Mater. **12**, 410 (2013).

[56] V. Sharma and G.S. Okram, Phys. B Condens. Matter **600**, 412453 (2020).

[57] Q. Du, M. Abeykoon, Y. Liu, G. Kotliar, and C. Petrovic, Phys. Rev. Lett. **123**, 076602 (2019).

[58] D. Byeon, R. Sobota, K. Delime-Codrin, S. Choi, K. Hirata, M. Adachi, M. Kiyama, T. Matsuura, Y. Yamamoto, M. Matsunami, and T. Takeuchi, Nat. Commun. **10**, 72 (2019).

[59] M. Cutler and N. F. Mott, Mott, Phys. Rev **181**, 1336 (1969).

[60] B. Xu and M. J. Verstraete, Phys. Rev. Lett. **112**, 196603 (2014).

[61] H. Fritz, Solid State Commun. **9**, 1813 (1971).

[62] D. K. Ko and C. B. Murray, ACS Nano **5**, 4810 (2011).